\documentclass[journal=jacsat,manuscript=article,layout=twocolumn]{achemso} % For two-column mode
\usepackage{chemformula} % Formula subscripts using \ch{}
\usepackage[T1]{fontenc} % Use modern font encodings
\usepackage[hidelinks]{hyperref}
\usepackage{xcolor}  % Enables highlighting (used for the revised version only)
\usepackage[export]{adjustbox} % Enables boxes around figures (used for the revised version only)
\graphicspath{ {./Figures/} }
\author{Anna Hartl}
\affiliation[LMX]{Center for Neutron and Muon Sciences, Paul Scherrer Institute, 5232 Villigen PSI, Switzerland}
\alsoaffiliation{Center for Photon Science, Paul Scherrer Institute, 5232 Villigen PSI, Switzerland}
\alsoaffiliation{Laboratory of Inorganic Chemistry, Department of Chemistry and Applied Biosciences, ETH Zürich, 8093 Zurich, Switzerland}

\author{Ján Minár}
\affiliation{New Technologies-Research Center, University of West Bohemia, 301 00 Plzeň, Czech Republic}

\author{Procopios Constantinou}
\affiliation{Center for Photon Science, Paul Scherrer Institute, 5232 Villigen PSI, Switzerland}

\author{Vladimir Roddatis}
\affiliation{GFZ German Research Centre for Geosciences, Telegrafenberg, 14473 Potsdam, Germany}

\author{Fatima Alarab}
\affiliation{Center for Photon Science, Paul Scherrer Institute, 5232 Villigen PSI, Switzerland}

\author{Arnold M. Müller}
\affiliation{Laboratory of Ion Beam Physics, Department of Physics, ETH Zürich, 8093 Zurich, Switzerland}

\author{Christof Vockenhuber}
\affiliation{Laboratory of Ion Beam Physics, Department of Physics, ETH Zürich, 8093 Zurich, Switzerland}

\author{Thorsten Schmitt}
\affiliation{Center for Photon Science, Paul Scherrer Institute, 5232 Villigen PSI, Switzerland}

\author{Daniele Pergolesi}
\affiliation[LMX]{Center for Neutron and Muon Sciences, Paul Scherrer Institute, 5232 Villigen PSI, Switzerland}

\author{Thomas Lippert}
\affiliation[LMX]{Center for Neutron and Muon Sciences, Paul Scherrer Institute, 5232 Villigen PSI, Switzerland}

\author{Vladimir N. Strocov}
\email{vladimir.strocov@psi.ch}
\affiliation{Center for Photon Science, Paul Scherrer Institute, 5232 Villigen PSI, Switzerland}

\author{Nick A. Shepelin}
\email{nikita.shepelin@psi.ch}
\affiliation[LMX]{Center for Neutron and Muon Sciences, Paul Scherrer Institute, 5232 Villigen PSI, Switzerland}
\title[]{Anionic disorder and its impact on the surface electronic structure of oxynitride photoactive semiconductors}
\begin{document}
%%%%%%%%%%%%%%%%%%%%%%%%%%%%%%%%%%%%%%%%%%%%%%%%%%%%%%%%%%%%%%%%%%%%%
\newpage
\begin{abstract}
The conversion of solar energy into chemical energy, stored in the form of hydrogen, bears enormous potential as a sustainable fuel for powering emerging technologies. Photoactive oxynitrides are promising materials for splitting water into molecular oxygen and hydrogen. However, one of the issues limiting widespread commercial use of oxynitrides is the degradation during operation. While recent studies have shown the loss of nitrogen, its relation to the reduced efficiency has not been directly and systematically addressed with experiments. In this study, we demonstrate the impact of the anionic stoichiometry of \ch{BaTaO_xN_y} on its electronic structure and functional properties. Through experimental ion scattering, electron microscopy, and photoelectron spectroscopy investigations, we determine the anionic composition ranging from the bulk towards the surface of \ch{BaTaO_xN_y} thin films. This further serves as input for band structure computations modeling the substitutional disorder of the anion sites. Combining our experimental and computational approaches, we reveal the depth-dependent elemental composition of oxynitride films, resulting in downward band bending and the loss of semiconducting character towards the surface. Extending beyond idealized systems, we demonstrate the relation between the electronic properties of real oxynitride photoanodes and their performance, providing guidelines for engineering highly efficient photoelectrodes and photocatalysts for clean hydrogen production.
\end{abstract}

%%%%%%%%%%%%%%%%%%%%%%%%%%%%%%%%%%%%%%%%%%%%%%%%%%%%%%%%%%%%%%%%%%%%%
\section{\label{sec:Introduction}Introduction}
%\section{\label{sec:Introduction}}

Solar-driven water splitting converts light into chemical energy, which is consequently stored in the form of \ch{O_2} and \ch{H_2} \cite{FujishimaHonda1972}. This sustainable generation of clean hydrogen fuel bears significant potential for non-fossil energy production in the future \cite{Khaselev1998,Crabtree2004,Qiu2022}. However, the efficiency of the involved photoactive materials must be improved for large-scale use. Indeed, the main challenge in the water splitting process is the production of molecular oxygen as part of the so-called oxygen evolution reaction (OER), whereby the efficiency is often hindered by high overpotentials and ill-defined reaction kinetics \cite{Seh2017}. During the past two decades, research in the fields of photocatalysis and photoelectrochemistry (PEC) has explored oxynitride materials that have demonstrated notable suitability for water splitting applications \cite{Maeda2007,Takata2020}.\

Oxynitrides are characterized by a mixed occupation of anionic sites by oxygen and nitrogen, bearing benefits for their use as energy materials in the following ways: I) increasing the nitrogen content reduces the band gap of these n-type semiconductors, allowing them to absorb light in the visible wavelength range \cite{Maeda2007}; II) the introduction of N 2\textit{p} bands close to the valence band maximum shifts the band edges in favor of the potentials necessary to drive the water splitting half-reactions \cite{Balaz2013}; and III) the changes in the electronic band dispersion, for example a modified curvature of the bands close to the Fermi level, can tune the kinetic properties of the charge carriers; this eventually affects the initial charge carrier generation and transport in the semiconductor, both impacting the surface kinetics of the water splitting process \cite{Mills2017,Qiu2022}.\

These arguments are favorable for the photo-applications of oxynitrides; however, under oxidative conditions of the OER, oxynitrides evince instability and are likely to lose nitrogen \cite{Chen2011}. Nitrogen loss in oxynitrides has been directly observed by means of x-ray photoelectron spectroscopy (XPS) \cite{Abe2010,Pichler2017_2}, ion scattering spectrometry and transmission electron microscopy (TEM) \cite{PourmandTherani2022} as well as by in situ gas chromatography \cite{Kasahara2002,Takata2007,Black2017,Luo2019} as direct evidence of the nitrogen evolution reaction running in parallel with the OER \cite{Higashi2013,Ouhibi2019}. Notably, no prior study has linked the loss of nitrogen in oxynitrides during PEC to the changes in their electronic properties, which are closely related to the water splitting efficiency and are therefore immensely more informative in the design of better performing materials. Certainly, it is the skeleton composed of metal-anion bonds that determines the rich physical and chemical properties of perovskite-type oxides \cite{Hwang2017,Strocov2022,Plumb2017} and oxynitrides. We therefore aim to understand the impact of the anionic composition of \ch{ABO_2N}-type perovskite oxynitrides, where \ch{A} and \ch{B} represent the cations in the general formula, on their electronic structure and subsequently correlate it to their water splitting performance.\

We consider the \ch{BaTaO_xN_y} (BTON) oxynitride (with \ch{x}=2 and \ch{y}=1 in the case of perfect stoichiometry) as a good candidate material due to a relatively small experimental band gap of $\sim$1.9~eV \cite{Haydous2019,Maeda2007} and suitable band edges for the production of oxygen (as well as hydrogen)\cite{Balaz2013,Luo2019}. Consequently, BTON has been extensively investigated and characterized in the form of powders \cite{Ueda2015,Hojamberdiev2021} and thin films \cite{Haydous2019} in the context of photoelectrochemistry during the past two decades \cite{Maeda2007}. Thin film samples offer the advantage of a defined surface eliminating morphological effects \cite{Haydous2019} and allowing to study the intrinsic properties of the material. Thus, our work deals with thin film samples of epitaxially-grown BTON with a (001) surface.\

By combining transmission electron microscopy (TEM) and ion scattering techniques with photoemission spectroscopy (PES), we have elucidated the anionic composition of BTON(001) thin films as a function of distance from the surface. Band structure computational modeling of the static disorder of anionic sites in conjunction with our PES analysis show that deviations of the anionic content from ideal stoichiometry have detrimental consequences for the surface band structure of the oxynitride semiconductor. Correlated with the depth-dependent composition, our BTON samples evince downward band bending towards the surface, where band inversion leads to the loss of semiconducting properties. Since this band bending effect extends over the majority of the expected space charge region, we argue that subtle changes in the anionic composition must have significant effects on the charge generation and charge transport during the OER, overall lowering the efficiency of BTON oxynitride photoanodes.\

Our study reveals surface metallicity of BTON oxynitrides and calls for more detailed characterization of oxynitrides in terms of their composition and optoelectronic response in future research. We also suggest how the band structure of oxynitrides near the surface can be tuned to match the requirements for the reaction kinetics and the potential landscape that drive the OER.

\section{\label{sec:Results}Results} 

\subsection{Thin film growth and characterization}
\begin{figure*}[h!]
  \includegraphics[width=\textwidth]{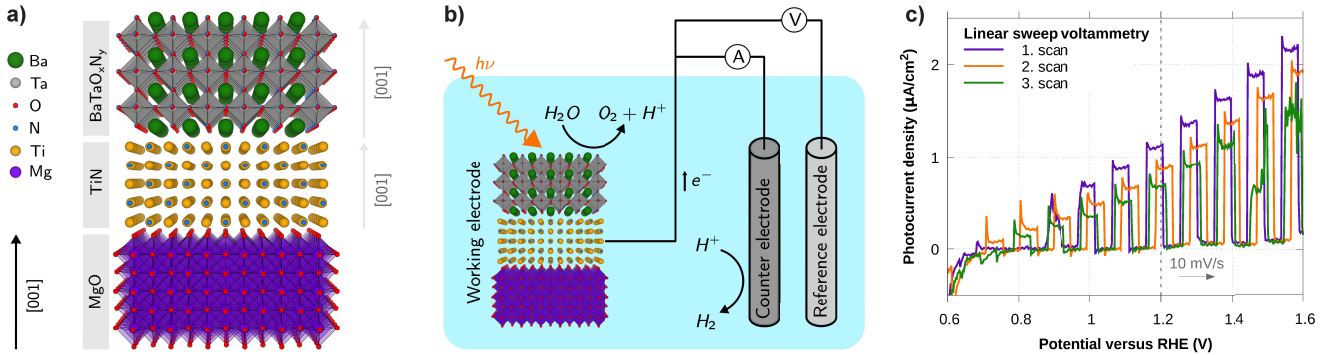}
  \caption{Oxynitride thin films and their photoelectrochemical operation. a) Structural model of the \ch{BaTaO_xN_y} thin film system. b) Schematic of the three-electrode configuration for photon($h\nu$)-induced water splitting, as explained in the~\ref{sec:Results}{Results} section. c) Three consecutive linear sweep voltammetric scans obtained from a BTON working electrode.}
  \label{Figure1} 
\end{figure*}

Thin film samples were fabricated by pulsed laser deposition and consist of an insulating MgO(001) substrate, a conducting TiN buffer layer acting as the current collector with a thickness of approximately 10~nm, and an approximately 116~nm thick BTON oxynitride layer, which was epitaxially grown with the [001] crystallographic direction out-of-plane, as shown in Figure~\ref{Figure1}a). BTON crystallizes in the space group $Pm\bar{3}m$ (No. 221) and shows a simple cubic lattice, as well as a simple cubic Brillouin zone in momentum space. The out-of-plane growth direction was confirmed by x-ray diffraction (see section~S1 in the Supporting Information (SI)) and transmission electron microscopy (TEM). Thereby, high-angle annular dark-field (HAADF) images show the cross section of the multilayer system and its crystal quality in Figure~\ref{Figure2}a-b) and Figure~S5 in the SI.\

\subsection{Photoelectrochemical performance}
Figure~\ref{Figure1}b) illustrates how the BTON films work as photoanodes for the OER. The corresponding photoelectrochemical setup is briefly described below, with further experimental details provided in the~\ref{sec:Methods}{Methods} section. The PEC testing was performed in the three-electrode configuration, consisting of the working electrode, which is the BTON film system, the counter electrode, and the reference electrode, all immersed in an electrolyte and connected to a potentiostat. The potentiostat sets the potential at the working electrode relative to the reference electrode (indicated by "V" in Figure~\ref{Figure1}b). Under solar radiation ("$h\nu$"), the generation of electron-hole pairs initializes water splitting. With applied positive bias, the generated electrons ($e^-$) in the oxynitride layer travel towards the TiN layer and further towards the counter electrode (photocathode), where the hydrogen evolution reaction takes place; simultaneously, the generated holes reach the oxynitride surface, where they drive the reactions of the OER and \ch{O_2} is produced \cite{Seh2017}. Proportional to the production of molecular oxygen during water splitting, the photocurrent density of the photoanode can be measured to determine its efficiency (marked by "A" in Figure~\ref{Figure1}b)).\

Electrochemical testing of BTON photoanodes was conducted under chopped irradiation, collecting the photoresponse of BTON thin films in terms of linear sweep voltammetry. The measured photocurrent density reached an average stabilized photocurrent of 1.12~µAcm$^{-2}$ at 1.2~V versus reversible hydrogen electrode (RHE) for the first scan in Figure~\ref{Figure1}c) and decreased gradually: it was diminished by $\sim$39\% in the third scan and resulted in an average stabilized photocurrent density of 0.47(7)~µAcm$^{-2}$ after the fifth scan until the last (ninth) scan. These results highlight the efficiency loss commonly observed for oxynitride photoanodes, and are in line with reported PEC experiments using \ch{BaTaO_xN_y} \cite{HaydousThesis}, \ch{SrTaO_xN_y} \cite{Lawley2021}, \ch{LaTiO_xN_y} \cite{Burns2020,Pichler2017_2,Lawley2020}, and \ch{CaNbO_xN_y} \cite{Haydous2021} thin films.\

\subsection{Elemental composition}
\begin{figure*}[h!]
  \includegraphics[width=\textwidth]{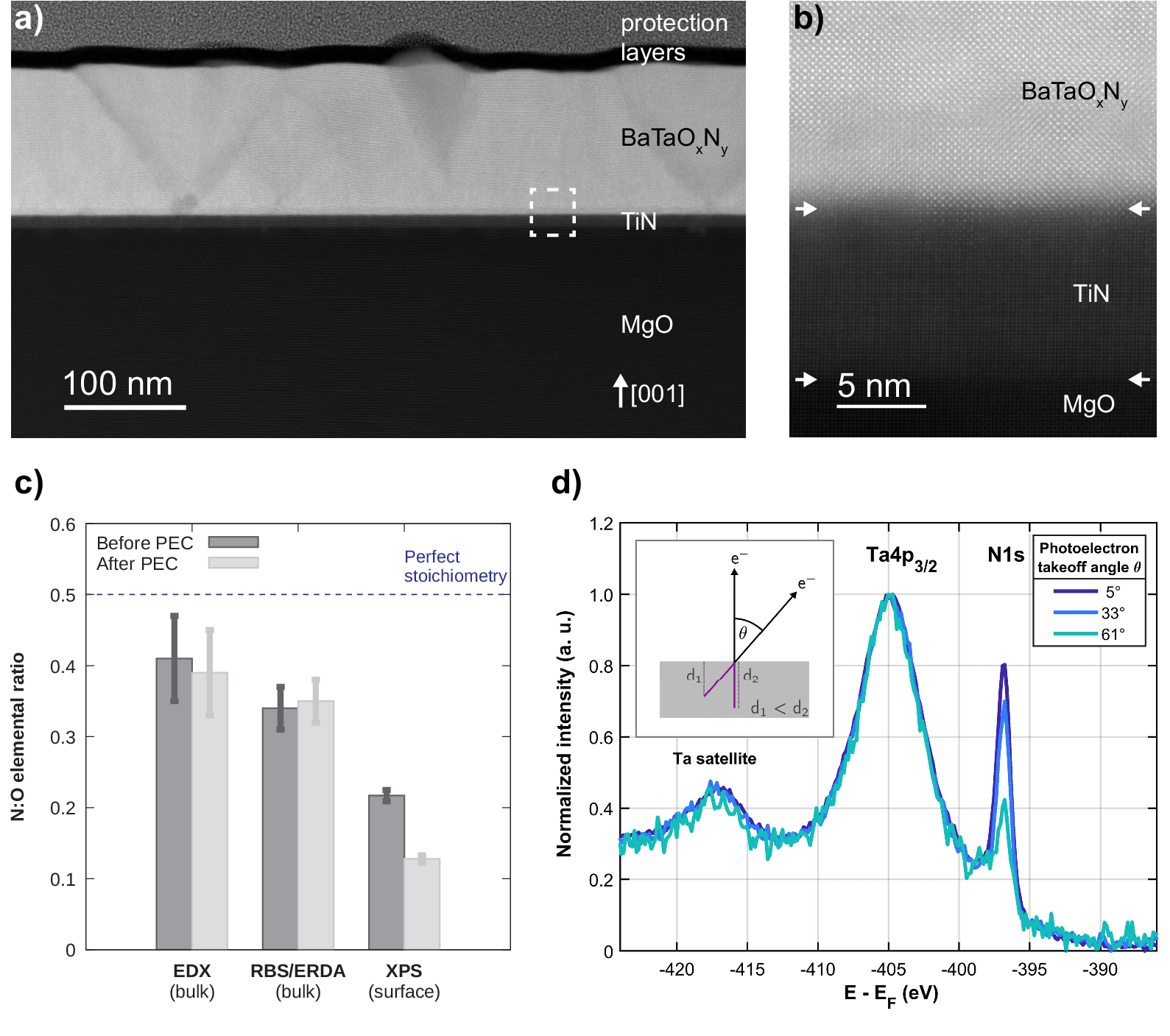}
  \caption{Structural and chemical analysis of the fabricated \ch{BaTaO_xN_y} thin film. a) Low magnification HAADF image of \ch{BaTaO_xN_y} before photoelectrochemistry (PEC) experiments. The protection layers were deposited for the TEM analysis only (see~\ref{sec:Methods}{Methods} section). b) Close-up section of the area denoted by the dashed box in a) indicating the TiN-MgO and \ch{BaTaO_xN_y}-TiN epitaxial interfaces with white arrows. c) Elemental nitrogen-to-oxygen ratio (N:O) obtained from the oxynitride thin film before and after PEC using complementary techniques. d) Angle-dependent XPS illustrating decreasing nitrogen abundance towards the surface (towards larger photoelectron takeoff angles).}
  \label{Figure2} 
\end{figure*}

The elemental composition of BTON before and after the electrochemical application was tracked by complementary techniques (see Figure~\ref{Figure2}c)), which are either bulk- or surface-sensitive. Ion scattering techniques such as Rutherford backscattering (RBS) and heavy-ion elastic recoil detection analysis (ERDA) can give accurate values for the bulk elemental composition of thin films \cite{Dobeli2005}. However, these techniques suffer from poor depth resolution when approaching the nm range. By using energy-dispersive x-ray (EDX) analysis, TEM allows to identify the bulk as well as the surface composition, even obtaining information about the N:O ratio for the top atomic layers close to the limit of the instrumental resolution, as represented by the large error bars in the SI Figure~S4. The most reliable technique for surface-sensitive studies is XPS. In the soft x-ray regime, an average probing depth of about 3-6~nm is obtained for  perfectly stoichiometric \ch{BaTaO_2N} using the TPP‐2M formalism of the inelastic mean free path $\lambda _{\text{IMFP}}$ of photoelectrons \cite{Tanuma1987,Shinotsuka2018} and the average probing depth being equal to $3\lambda _{\text{IMFP}}$. The depth sensitivity can be tuned by varying the kinetic energy or the take-off angle of emitted photoelectrons. The former can be adjusted via the incident photon energy ($h\nu$) and the latter is graphically visualized in the inset of Figure~\ref{Figure2}d).\

While x-ray diffraction and TEM indicate no significant structural changes (SI Figure~S1-S5) of thin films before and after the PEC application, a change in the N:O ratio becomes evident for the BTON surface, as shown in Figure~\ref{Figure2}c). Qualitatively, this trend is reproduced by angle-dependent XPS (Figure~\ref{Figure2}d)).\

\subsection{Anionic disorder and its theoretical implementation}
Modeling atomic disorder can be challenging \cite{Minar2014} and several approaches have been developed over the past decades to implement static disorder in the frame of multiple-scattering theory \cite{Bellaiche2000,Bansil1979,Soven1967}. For oxynitrides, recent computational approaches have suggested atomic substitution of oxygen by nitrogen in an ordered fashion. This means that the nitrogen atoms preferentially occupy anionic sites resulting in \textit{cis}-type and \textit{trans}-type anion configurations \cite{Lan2021,Ouhbi2018,GarciaCastro2022,Page2007} or even more complex scenarios \cite{Camp2012}. Certainly, these local anionic arrangements could simplify the computational implementation of oxynitride systems. However, to the best of our knowledge, specific anion configurations have not been experimentally observed in thin films. We argue that local anion ordering cannot necessarily be applied to real samples with non-perfect stoichiometry and since no studies performed to date have successfully shown direct experimental evidence of such anionic ordering, we considered a random distribution of anions. Using the SPR-KKR program package, we implemented such mixed occupancy sites in the frame of the coherent potential approximation (CPA) suitable for systems showing strong static disorder \cite{Niklaus2019}. Herein, the anionic sites could either host O, N, or vacancies V. Although we observe signs of a Ba-enriched surface in our XPS evaluation (see discussion in the SI section~S6), for a first approximation analysis applicable to our soft x-ray probing depth, we consider a constant Ba:Ta ratio of 1 throughout the entire film corroborated by our TEM and RBS/ERDA results.\

In the following paragraph, we explain the determination of BTON stoichiometries including vacancy content, which we label as \ch{V^O_{x'}} and \ch{V^N_{y'}} in its stoichiometric formula, resulting in \ch{BaTaO_{2-x'}N_{1-y'} V^O_{x'} V^N_{y'}}. Here, \ch{V^O_{x'}} represents the fraction of divalent oxygen removed from the crystal lattice. Likewise, \ch{V^N_{y'}} is the fraction of removed nitrogen and is trivalent. Taking the valences of the anions and vacancies into consideration, their stoichiometric coefficients must sum up to -7 to conserve charge neutrality. We note that not all compositions resulting from this formula are stable and can be realized experimentally.\

Nitrogen tends to leave the atomic lattice near the surface. Our angle-dependent XPS studies captured this loss of nitrogen in samples before the PEC and increased loss due to the nitrogen evolution reaction in samples after the PEC (see Figure~\ref{Figure2}c)). It seems that oxynitrides are characterized by a nitrogen-depleted surface, regardless of the surface treatment, which is mentioned in the SI section~S8. This can be understood as follows: the lower bonding strength of the cations with nitrogen in comparison to oxygen leads to a band structure with N 2\textit{p} states hybridizing at higher energies than the O 2\textit{p} states\cite{Maeda2007}, as we will discuss later. Thus, we expect that most vacancies stem from nitrogen. In addition, it has been reported that vacancies can also stem from oxygen and even complex interplays can exist \cite{Xiong2006,Baumgarten2021}. Because there are several possible routes for the evolution of the anionic stoichiometry in BTON samples, further approximations to constrain the unknown coefficients are needed.\ 

We obtain \ch{O_{2-x'}} and \ch{V^O_{x'}} based on the O:Ta ratio from XPS (see Table~S2 in the SI). Further, the N:O ratio from XPS allows us to estimate \ch{N_{1-y'}} and \ch{V^N_{y'}}. The resulting stoichiometries for our BTON sample are found to be \ch{BaTaO_{1.85(5)}N_{0.403(18)}V^O_{0.147(4)}V^N_{0.597(27)}} before PEC and \ch{BaTaO_{1.86(5)}N_{0.239(11)}V^O_{0.138(4)}V^N_{0.76(3)}} after PEC. For the sake of readability, these are referred to hereafter as \ch{BaTaO_{1.85}N_{0.40}V_{0.75}} and \ch{BaTaO_{1.86}N_{0.24}V_{0.90}}, respectively.\

First, we compared the electronic band structure in terms of the Bloch spectral function (BSF) representing the oxide-phase (\ch{BaTaO_{3.5}}) and the ideal oxynitride (\ch{BaTaO_2N}) in the SI Figure~S6a) and b), respectively. While the oxide shows a well-defined valence band and a band gap of 3.5~eV, the band gap of the stoichiometric oxynitride is reduced to $\sim$1.7~eV. Furthermore, the valence band of the oxynitride is broadened over the momentum- and energy-range due to the inherent atomic disorder. This band smearing is most pronounced within the top 2~eV of the valence band and coincides with the energy range where the N 2\textit{p} states become apparent in representation of the density of states $n(E)$.\

It remains to be a question for future work, to elucidate how the broadening of the bands (the localization of valence electrons) is related to the initial charge carrier formation during PEC. The assignment of the Ta valence band states to Ta 5\textit{d} classifies this material as a charge-transfer insulator according to the Mott-Hubbard model \cite{Greiner2013}. This type of electronic system can be prone to the loss of lattice anions, as discussed by Gent et al. \cite{Gent2020} in terms of the oxygen redox of top valence band states and oxygen dimer formation in oxides.\

\begin{figure*}[h!]
  \includegraphics[width=\textwidth]{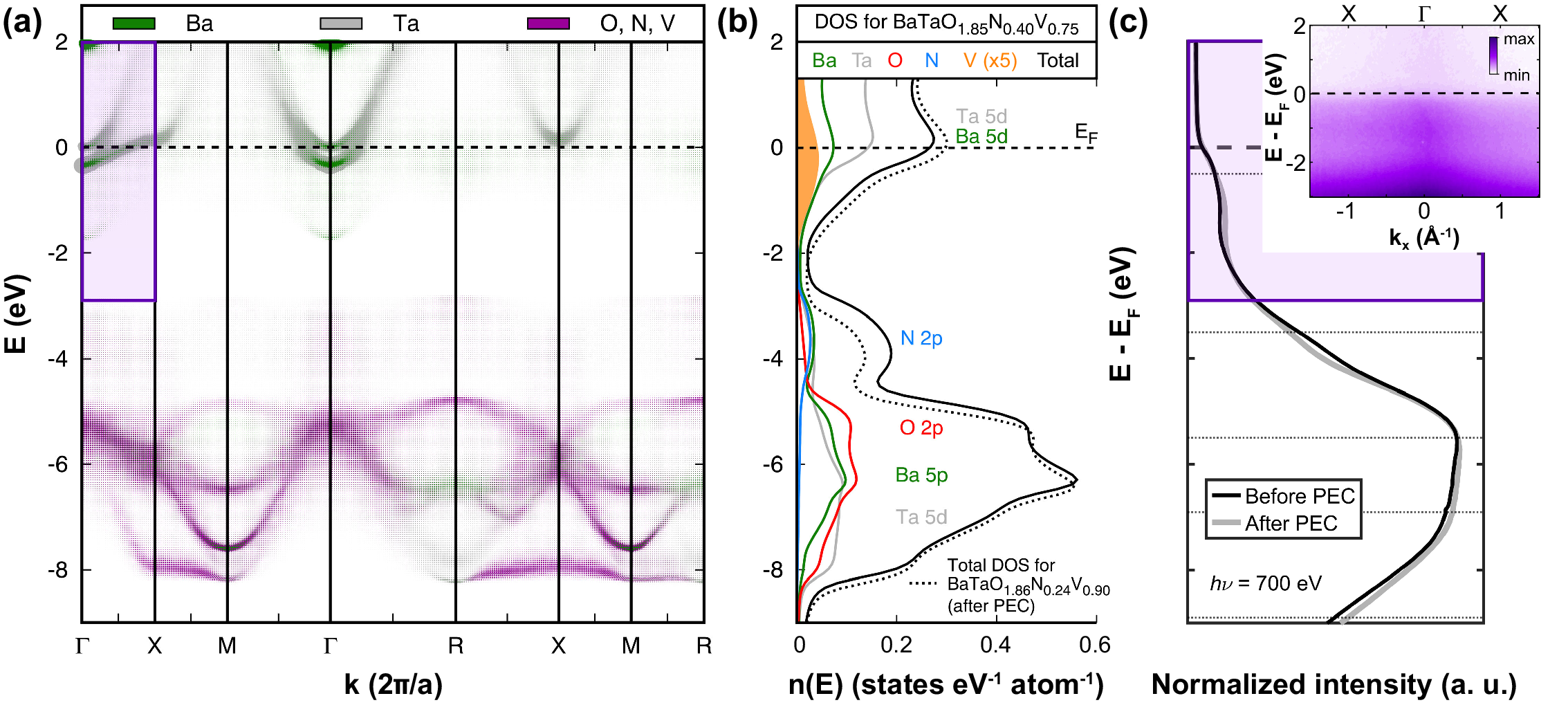}
  \caption{Computational and experimental features of the BTON band structure. a) Site-dependent Bloch spectral function of \ch{BaTaO_{1.85}N_{0.40}V_{0.75}} with typical features of an anion-deficient oxynitride shown over characteristic momentum paths. The purple box marks the band structure region that has been explored experimentally, which is displayed in panel c). b) The total and partial density of states for the same system highlighting the main orbital contributions to the valence band region. The total density of states for \ch{BaTaO_{1.86}N_{0.24}V_{0.90}} corresponding to the composition after photoelectrochemistry (PEC) is represented by the black dotted line. c) Experimental momentum-integrated photoemission spectrum for BTON before and after PEC. The inset shows the momentum-resolved data for the sample before PEC.}
  \label{Figure3} 
\end{figure*}

To relate BSF computations to our real oxynitride samples, we implemented the determined oxygen, nitrogen, and vacancy content in our computations. Figure~\ref{Figure3} juxtaposes the calculated BSF and the experimental band structure data from photoelectron spectroscopy (PES).\

In comparison to perfectly stoichiometric \ch{BaTaO_2N}, the introduction of anion-derived vacancies in \ch{BaTaO_{1.85}N_{0.40}V_{0.75}} affects both the valence and conduction band regions in terms of increased band smearing (Figure~\ref{Figure3}a) and SI Figure~S6c)). The valence band states between -8 and -5~eV in Figure~\ref{Figure3}a) show pronounced broadening and flattening of bands (highlighted by the purple-colored anionic bands), indicating increased charge localization and enhanced effective hole masses, respectively. The latter must lower the mobility of holes in the oxynitride and impact its PEC performance \cite{Qiu2022}.\

The valence band region between -5 and -3~eV in the off-stoichiometric BTON is predominantly occupied by N 2\textit{p} hybridizing with the Ta 5\textit{d} and Ba 5\textit{p} states. In fact, these non-dispersive cation-derived states populate the band gap region near both band edges. These in-gap states are reproduced in our experiments, whereby we observe a populated "band gap" below the Fermi level ($E_F$) in angle-resolved PES data (inset in Figure~\ref{Figure3}c)). We expect that these Ta- and Ba-derived defect states appear due to the highly-disordered charge environment around the cations caused by the defective anionic sublattice.\

The vacancies present in \ch{BaTaO_{1.85}N_{0.40}V_{0.75}} also introduce new bands around the $\Gamma$ point, located at around -1.8~eV in Figure~\ref{Figure3}a) and highlighted in orange in Figure~\ref{Figure3}b). These bands were also accessed experimentally with PES (Figure~\ref{Figure3}c)) and are faintly discernible in the angle-resolved data (inset in Figure~\ref{Figure3}c)), centered around $k_x$=0~\textup{~\AA}$^{-1}$. The fact that those vacancy-derived bands are localized in momentum space, means that the corresponding electrons are delocalized in real space following the periodicity of the lattice. Again, we stress that the periodicity of the anionic sites was modeled using mixed occupancies and did not follow special local arrangements around the tantalum octahedra. The presence of these bands in the theoretical and experimental data supports our computational approach. Indeed, this agreement is an argument in favor of an in-plane random distribution of O, N, and V sites.\

In addition, the total density of states for the system based on the BTON composition after PEC (\ch{BaTaO_{1.86}N_{0.24}V_{0.90}}) was drawn into Figure~\ref{Figure3}b) by the dashed line, which demonstrates a decreased population of N 2\textit{p} states and an overall downward-shift of the bands by -0.13~eV focusing on the predominant O 2\textit{p} bands. This is in agreement with our experiments, where angle-integrated PES resolved a subtle $\sim$-0.15~eV global downward shift of the valence band for BTON after PEC in comparison to the sample before PEC (gray line in comparison to the black line in Figure~\ref{Figure3}c)) as well as decreasing spectral weight for features located at the top 2~eV of the valence band, $E-E_F$=-3.5~eV, and $E-E_F$=-6.9~eV. In total, these changes are minor as can be expected from the valence band dominated by the spectral weight of barium (prevalence of the Ba 5\textit{p} cross section \cite{YehLindau1985}). Due to the absence of indicative shifts in energy for certain bands, we excluded notable changes in the degree of hybridization \cite{Heymann2022} between tantalum and the anions for this particular probing depth of $\sim$4~nm. This agrees with conclusions about the coordination environment for tantalum in \ch{SrTaO_xN_y} after PEC \cite{Lawley2021}.\ 

Overall, anionic vacancies in BTON create excess electrons that occupy the in-gap states, inducing metallic behavior. This becomes apparent by the upward shift of $E_F$ (or the downward shift of the band structure with respect to $E_F$) in our computations (Figure~\ref{Figure3}b)) and experiments (Figure~\ref{Figure3}c)). Figure~\ref{Figure3} already indicates the changes of the band structure as a function of the anionic composition from a sample before and after PEC. For a better understanding of the electronic properties, we tracked the shifts of the band structure with respect to the $E_F$ for the BTON sample after PEC by means of XPS, described in the next section.\

\subsection{Band bending at the surface region}
In this section, we relate the anionic composition of BTON to the alignment of its original band edges as a function of depth using XPS. The extraction of the depth-dependent potential near the surface is in general non-trivial for PES in the valence band region \cite{Schuwalow2021,Lev2015}. This holds especially in our case, where a depth-specific anionic composition additionally modulates the states near $E_F$. Consequently, valence band leading edge fits must lead to erroneous results \cite{Schuwalow2021} for BTON samples. Thus, we accessed information about the surface band alignment of the oxynitride after PEC with $h\nu$-dependent XPS on the Ta 4\textit{f} core level. We assume that a global shift in energy of the BTON band structure translates to its core levels. For the core level region, we can also neglect considerable changes in the matrix elements, as opposed to the matrix-element effects for the valence band region. We describe the fitting approach of the core levels in greater detail in the SI section~S5.\

\begin{figure*}[h!]
  \includegraphics[width=\textwidth]{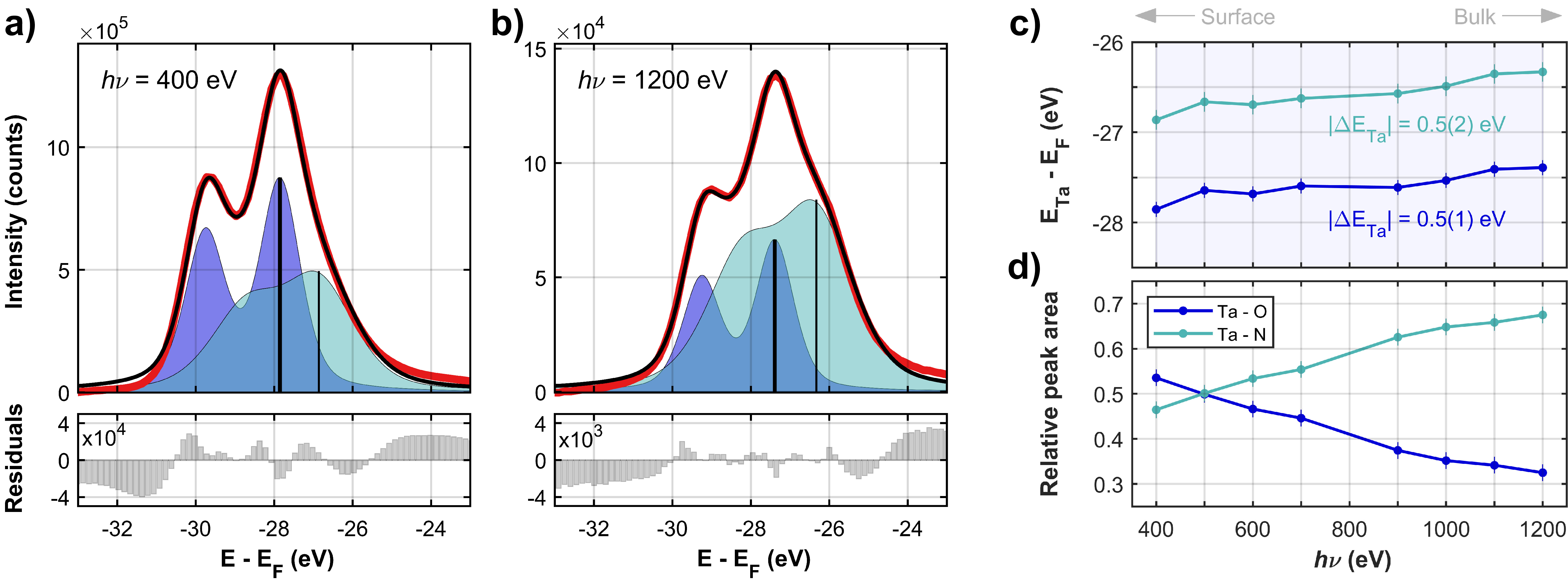}
  \caption{$h\nu$-dependent XPS with the Ta 4\textit{f} core level of \ch{BaTaO_xN_y} after PEC. a) The Ta 4\textit{f} core level characterized by the oxygen- (dark blue) and nitrogen- (light blue) derived doublet accessed with $h\nu$=400~eV corresponding to most surface-sensitive probing depth. The red line is the data, the black one is the fit. b) The Ta 4\textit{f} core level spectrum for $h\nu$=1200~eV corresponding to most bulk-sensitive probing depth. c) Extracted energies for both core level doublets as a function of the $h\nu$. |$\Delta E_{Ta}$| gives the absolute change in energy for a core level over the used $h\nu$ range. d) The relative contribution of each doublet to the fitted intensity of the Ta 4\textit{f} core level as a function of the $h\nu$.}
  \label{Figure4} 
\end{figure*}

The increase of the incident photon energy leads to an increase in the kinetic energy of the photoelectrons, allowing greater depths of the samples to be probed \cite{Strocov2013}. This is schematically shown in Figure~\ref{Figure4}a-b). However, since the core level signal for a given $h\nu$ is a superposition of signals from all probed atomic layers following Beer-Lambert law \cite{Huang2019}, the average probing depth cannot be set equal to the actual depth. This complicates the extraction of the depth-dependent band alignment, represented by the Ta 4\textit{f} core level energies. For a measurement at a certain $h\nu$, the depth can be estimated when the electrostatic potential is known \cite{Schuwalow2021,Huang2019}. Because no exact descriptions of potential profiles for oxynitride semiconductors exist, up to our knowledge, we have established a description of the depth-dependent potential profile and the band profile in the SI section~S9, based on nonuniform doping as a result of the change in anionic composition within the first 10~nm of BTON.\

The Ta 4\textit{f} core level consists of two doublets originating from the Ta-O (doublet at lower energies) and the Ta-N bonding environment. (See SI section~S7 for a discussion about the Ta 4\textit{f} core level assignment.) The fitted energy for each core level ($E_{Ta}-E_F$), exemplified by the vertical black lines in Figure~\ref{Figure4}a) and~b), as a function of $h\nu$ is displayed in Figure~\ref{Figure4}c). Here, a decrease for the Ta-O- and the Ta-N-derived core level energies towards more surface-sensitive $h\nu$ becomes apparent. This change in $E_{Ta}$ across the probed $h\nu$ range is in the order of 0.5~eV for both core levels. In Figure~\ref{Figure4}d), we show the integrated peak areas for the Ta-O and the Ta-N doublets. The almost linear decrease of the nitrogen-related peak area (Figure~\ref{Figure4}d)) towards more surface-sensitive $h\nu$ indicates the loss of nitrogen. This is in agreement with the angle-dependent XPS study shown in Figure~\ref{Figure2}d).\

\begin{figure}[h!]
  \includegraphics[width=\columnwidth]{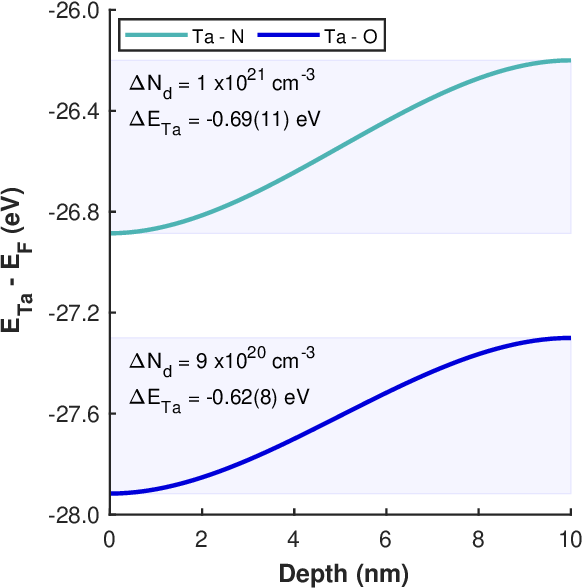}
  \caption{Representation of the Ta 4\textit{f} core level of \ch{BaTaO_xN_y} after photoelectrochemistry as a function of depth according to the model for nonuniform doping. The color-coded band profiles for both core level doublets are based on our model using the best found value for $\Delta N_d$ (the linear change in dopant concentration across the space charge region). The resulting value for the absolute band bending over the space charge region is also given with $\Delta E_{Ta}$.}
  \label{Figure5} 
\end{figure}

Going from the bulk towards the surface of the BTON films, a gradual change in the anionic composition, exemplified by a decrease of the nitrogen and increase of the vacancy content, induces n-doping, which is accompanied by downward band bending. We addressed this type of non-constant doping across the space charge region in terms of linear n-doping, implemented in our model for a nonuniformly doped semiconductor in SI section~S9. This nonuniform doping results in an unusual band profile for depths up to 3~nm from the surface (Figure~\ref{Figure5})) in our obtained model. Applied to both Ta 4\textit{f} core levels, the total band bending (or the band offset between the oxynitride and the vacuum region) is in the order of -0.6~eV.\

\section{\label{sec:Discussion}Discussion}
We quantified the N:O ratio in \ch{BaTaO_xN_y} oxynitride thin films before and after the described PEC application (ex-situ) with complementary techniques. While no notable changes could be detected for the bulk composition using EDX and RBS/ERDA, XPS with $h\nu$=1100~eV allowed to track the decrease in the N:O ratio by $\sim$41\% for BTON after the PEC. Our core level analysis also suggests that part of the anionic vacancies in BTON samples are replaced by oxygen during the OER and that the higher electronegativity of oxygen ensures the pentavalent oxidation state of tantalum. Another indication that this happens in BTON samples is the prevalence of the Ta-O contribution to the Ta 4\textit{f} core level towards the surface (see Figure~\ref{Figure4}d)).\

Based on the XPS results, the anionic composition of BTON samples was determined as an integrand value over the probing depth for the actual $h\nu$ and implemented in our band structure calculations. The computational results are in qualitative agreement with the spectral features in angle-integrated PES. As discussed above, the BTON band structure is characterized by features arising from the static anionic disorder. Most notably, the loss of N 2\textit{p} valence band states, exposing the cation (defect) states, is accompanied by the shift of $E_F$.\

$h\nu$-dependent XPS of the Ta 4\textit{f} core level for the BTON sample after PEC evinced a chemical gradient in terms of a decreasing N:O ratio towards the surface and a shift of the fitted energies across the experimentally accessible probing depth. Based on these compositional changes, the established model description for nonuniformly doped BTON enabled a band profile for the Ta 4\textit{f} core level with depth. The observed downward band bending in the order of -0.6~eV can be expected for systems after surface treatment \cite{Caban-Acevedo2014,Kim2001}.\

Based on these observations for BTON, we now address how the electronic structure relates to the semiconductor physics and its PEC performance. The vacancies in a real and defective BTON system induce band structure characteristics and have detrimental influence on the electronic and functional properties:\

I) Anionic vacancies in the oxynitride create excess electrons that lead to charge accumulation in the n-type semiconductor. In a simple band diagram, this leads to downward band bending towards the surface. According to the used model for nonuniformly-doped oxynitrides, one can expect a built-in field leading to band bending under equilibrium conditions, which is opposed to the customary representation of oxynitrides under flat band conditions. Within the band bending region of our sample, the position of the Fermi level for our non-stoichiometric oxynitride (Figure~\ref{Figure3}) is expected to be above the conduction band minimum of a stoichiometric oxynitride as displayed in Figure~S6b).\cite{Balaz2013}. This is characteristic for a degenerate n-type semiconductor and agrees with the relatively high dopant concentrations from the band bending analysis mentioned in Figure~\ref{Figure5}). The resulting band inversion leads to surface metallicity and a decrease in the photocurrent. This type of band alignment is undesired for the photoanodes and must be overcome during PEC. Conversely, upward band bending facilitates the transport of holes to the surface, where they "feed" the charge transfer processes for the OER. II) Emerging in-gap states must considerably contribute to the charge carrier recombination, further decreasing the photocurrent. III) Surpassing a critical density of in-gap states, the oxynitride semiconductor turns into a metal and possibly loses its function as a photoactive material.\

Notably, BTON shows only little modification in terms of its crystal structure across the PEC. In additional experiments (not reported), we were able to grow epitaxial BTON films even for low nitrogen contents (with N:O < 1\% according to XPS with $h\nu$=1100~eV). The fact that the simple cubic lattice persists at strongly-defective anionic compositions highlights the exceptional structural stability of the lattice. Likely, further oxidation of the compound in the frame of the OER must introduce lattice instability exceeding strain effects, as discussed in the SI section~S1, and structural decomposition. The spectral weight of the in-gap states as seen in PES could therefore be a precursor state of BTON before its lattice breaks down. Indeed, we hypothesize that the structural robustness of BTON prevents charge compensation and promotes in-gap states. To overcome this intrinsic problem, we share strategies to overcome charge imbalances from a materials design perspective in the following paragraph.\

In general, the electronic properties for idealized photoactive oxynitrides, such as those found in computational analyses, can only be achieved in a perfectly-stoichiometric structure. In real systems, however, anionic vacancies will most likely be present. To compensate this charge imbalance and thereby improve the electronic properties of oxynitride photoanodes, we suggest following routes: I) intrinsic p-doping (instead of n-doping \cite{Higashi2015} by A and B site cations, resulting in a monovalent cation (such as \ch{K^+}, \ch{Na^+}, \ch{Li^+}, or \ch{Ag^+}) for the A site or a tetravalent cation (such as \ch{Ti^{4+}}, \ch{Zr^{4+}}, \ch{Hf^{4+}}, or \ch{Sn^{4+}}) for the B site \cite{Hojamberdiev2022}, in the case of BTON; II) p-doping by the fabrication of heterostructures including \ch{A(O,N)_1} and \ch{B(O,N)_2} layers, where A and B stand for the aforementioned cations in the case of BTON; and/or III) employing oxynitrides with more flexible structural frameworks. For the latter, octahedral tilting and cationic displacements could counterbalance excess electrons in the semiconductor. Herein, proposed \ch{ABO_2N}-type compounds with orthogonal conventional bases, such as orthorhombic, tetragonal, and cubic crystal systems, and octahedral tilting in one, two, or even three crystallographic directions could be used. Common examples for these three tilt systems are \ch{SrTaO_2N}, \ch{SrNbO_2N} with space group $I4/mcm$ (No.~140), \ch{LaTaO_2N}, \ch{LaTiO_2N} with $Imma$ (No.~74), and \ch{CaTaO_2N}, \ch{CaNbO_2N}, \ch{CaMoO_2N}, \ch{CeNbO_2N}, \ch{LaZrO_2N}, and \ch{NdTiO_2N} with $Pnma$ (No.~62), respectively. However, the tilting can increase the phonon-mediated charge transport, thereby inhibiting hole transport in the semiconductor. Thus, one must be careful optimizing the system.\

In addition to the suggested improvements for the photoresponse of oxynitrides, one needs to boost the catalytic properties. Species possessing higher OER activity such as transition metal oxides should be introduced, for example as a top co-catalyst \cite{Ueda2015,Wang2021,Lawley2020}. Top metal oxides could further aid in the passivation of the surface \cite{Gamelin2012,Hoang2011} and counteract the preferred oxidation of BTON. We expect the above mentioned points to be transferable to other commonly-studied oxynitrides, such as \ch{SrTaO_xN_y}, \ch{LaTiO_xN_y}, and \ch{CaNbO_xN_y}.\

\section{\label{sec:Conclusion}Conclusion}
We have elucidated the characteristic band structures of real oxynitride thin films with defective anionic compositions and contributed to the depth-dependent and in-depth understanding of the oxynitride band structures and their evolution across the PEC.\

For oxynitrides, we expect that the OER runs in parallel with the nitrogen evolution reaction involving the loss of lattice nitrogen. This reaction has been detected in past experiments with powders \cite{Kasahara2002,Takata2007,Black2017,Luo2019} and films \cite{Lawley2020,Lawley2021}, but was usually considered to be of negligible significance. Our investigation on the electronic structure of BTON samples, however, proved the opposite and revealed the detrimental consequences of nitrogen loss and anionic disorder of oxynitrides in general: anionic deficiencies can even lead to the loss of the semiconductor properties at the surface, where inversion of the n-type semiconductor turns the compound into a metal and must consequently hinder the electron-hole pair formation. In such a case, the desired characteristics of photoactive oxynitrides are lost.\

Finally, the electronic properties of an oxynitride semiconductor are inherently tied to its anionic stoichiometry and need to be addressed before "trial-and-error" engineering of oxynitride films and powders. Our work has investigated the very basic properties of BTON, and therefore forms a basis for tuning the electronic structure of oxynitride photoelectrodes through future studies with angle-resolved PES.\

\section{\label{sec:Methods}Methods}
\subsection{Pulsed laser deposition of oxynitride thin films}
For each pulsed laser deposition, a 10x10~mm$^2$ large MgO(001) substrate (CrysTec GmbH) was used. To facilitate several sample characterizations for the same sample batch, each MgO substrate was cut into four equally-sized pieces. All data discussed in this publication stems from a single sample batch. Only the angle-dependent XPS data originates from separate batch produced under the same experimental conditions.\

The cut 5x5~mm$^2$ substrates were sonicated for 15 minutes in acetone followed by 15 minute-long sonication in propan-2-ol before thermal annealing in a tube furnace at 1000~$^{\circ}$C for 12 hours. The heating ramp of the tube furnace was 5.8~$^{\circ}$C per minute and cooling proceeded in a passive manner.\

The substrates were mounted on a substrate holder with Ag paste (SPI Supplies, Structure Probe Inc.), which was cured in a stepwise fashion (50-120~$^{\circ}$C) for at least 45 minutes. The substrate holder was then introduced to the deposition chamber and gradually heated to 700~$^{\circ}$C. We note that sufficiently high temperatures at the substrate are crucial for crystal quality and nitrogen incorporation during deposition.\

Thin film systems were fabricated using a pulsed KrF excimer laser (LPX305iCC, Lambda Physik), with $\lambda$=248~nm and a measured fluence of 2.3~Jcm$^{-2}$, ablating material from a rod-shaped target that was positioned 50~mm away from the sample substrate. A TiN target was ablated for 30 minutes with a set laser frequency of 10~Hz while the pressure was kept at 2$\cdot$10$^{-6}$~mbar. This was followed by a 30-minute-long ablation of a \ch{Ba_5Ta_4O_15} target with frequency of 7~Hz. In this step, a nitrogen and ammonia gas mixture, with a molar \ch{NH_3}:\ch{N_2} ratio of 1:2/3, served as the background gas to create the reducing environment necessary for oxynitride production \cite{Marozau2011}. The total pressure in the chamber during deposition was 2$\cdot$10$^{-6}$~mbar.\

\subsection{Photoelectrochemical testing}
PEC measurements were executed in a three-electrode configuration consisting of a Ag/AgCl reference electrode (stored in saturated KCl), a Pt counter electrode, and the BTON working electrode attached to a \ch{TiN}-coated stainless steel wire. We used a custom-designed reactor cell \cite{Lawley2021} of polyether ether ketone (PEEK) with 190~µm-thick Mylar foil as the window. The cell was filled with 0.5~M NaOH aqueous electrolyte solution.\

A 50-500 W Xenon DC arc lamp (Oriel Instruments, Newport) with the 69907 Universal Arc Lamp Power Support (Newport) was employed as the light source. Hereby, a lens and XLP12 power detector system (Gentec) in conjunction with the PC interface and software Integra (Gentec-EO) helped to define incoming light of 100~mWcm$^{-2}$ (equal to 1~sun).\

Linear sweep voltammograms were collected at room temperature (25~$^{\circ}$C) between 0.6 and 1.6~V versus RHE with a scan rate of 10~mVs$^{-1}$ with the µAutolab type III/FRA2 (Metrohm Autolab) potentiostat using the software Nova. The initial (zeroth) potentiodynamic scan was executed under dark conditions (not shown in Figure~\ref{Figure1}c), followed by measurements under chopped illumination with a period of 3~seconds.\

The potential $E_{\text{Ag/AgCl}}$ measured against the reference electrode with $E_{\text{Ag/AgCl}}^0$ was converted to the potential versus RHE ($E_{\text{RHE}}$), as shown in Figure~\ref{Figure1}c, via the Nernst equation $E_{\text{RHE}}=E_{\text{Ag/AgCl}}^0+0.05916\cdot pH+E_{\text{Ag/AgCl}}$. Taking the pH-value of the electrolyte (pH$_{\text{0.5 M NaOH}}$=13.55) and the sample area in contact with the electrolyte into account, the current-potential curves as shown in Figure~\ref{Figure1}c) were obtained.\

For comparison, we enlist the values of the measured photocurrent normalized with respect to the sample surface at 1.2~V versus RHE potential, as marked by the dashed line in Figure~\ref{Figure1}c). We have performed the PEC experiments on three BTON samples grown under comparable deposition conditions, giving a mean value of 1.02(20)~µAcm$^{-2}$ for the first linear sweep voltammetric scan. While one of the samples did not give further response after the first cycle, the other two samples showed average values of 0.93(4) and 0.82(14)~µAcm$^{-2}$ for the second and third scan, respectively. We note that it was not possible to determine the sample surface exposed to the electrolyte in the sub-millimeter-regime. Herein, possible fluctuations of the meniscus at the electrolyte-sample-air interface must lead to non-negligible deviations. The photocurrent was also not iR corrected. For the quantification of the PEC performance, however, we propose extended tests and long-term cycling.\

The photocurrent density stabilizes after the fifth scan, where it reaches an average "stabilized" photocurrent density \cite{HaydousThesis,Pichler2017_2} (averaged over the sixth until the ninth scan) of 0.47(7)~µAcm$^{-2}$.\

The so-called "recombination peaks", which are the spikes at the onset of the photocurrent steps in Figure~\ref{Figure1}c), commonly observed for oxynitrides \cite{Ueda2015,Haydous2021,Lawley2020,Pichler2017_2,Lawley2021}, are more pronounced when the light is turned on than off. This asymmetry suggests that more complex mechanisms, besides the recombination of surface holes, are involved \cite{Peter2020}.\

\subsection{Transmission electron microscopy}
Protection layers were deposited on the BTON surfaces for electron microscopy (sample preparation and imaging): from the BTON film to the sample surface, the thin protective layers consist of $\sim$10~nm amorphous carbon (dark layer in~\ref{Figure2}a) and $\sim$100~nm Pt (thicker top gray layer in~\ref{Figure2}a)) as well as further $\sim$1600~nm Pt (not shown in~\ref{Figure2}a)) deposited with an electron beam and ion beam, respectively. TEM specimens were prepared via the focused ion beam lift-out technique using a Thermo Fisher Scientific Helios 4UC instrument operated at 30~kV, and final cleaning was performed at 5~kV and 2~kV. High-angle annular dark field (HAADF) images and EDX maps (see SI Figure~S3) were collected with a Themis Z 3.1 60-300 FEG Scanning Transmission Electron Microscope (S\/TEM) operated at 300 kV which was equipped with a Gatan Imaging Filter Continuum 1065 and a SuperX Energy Dispersive X-ray detector.\

\subsection{RBS/ERDA}
Rutherford backscattering (RBS) and heavy-ion elastic recoil detection analysis (ERDA) elemental analysis were carried out at the 1.7 MV Tandetron accelerator facility located at the Ion Beam Physics Laboratory at ETH Zurich, Switzerland. RBS was conducted with a 2~MeV $^4$He beam. For the detection of the backscattered ions, a Si PIN diode was used. The corresponding data was analyzed using the RUMP program \cite{Doolittle1986}. To detect the relatively light oxygen and nitrogen atoms in the sample, heavy-ion elastic recoil detection analysis (ERDA) with a 13~MeV $^{127}$I beam with 18° incident and exit angle, a time-of-flight spectrometer, and gas ionization detector was performed. The depth profiles were analyzed with the Potku software \cite{Arstila2014}.\

\subsection{PES}
All PES experiments were performed at the soft x-ray ARPES endstation \cite{Strocov2010} as part of the ADRESS beamline \cite{Strocov2013} at the Swiss Light Source, PSI, Switzerland. Measurements benefited from the high flux up to 10$^{13}$~photons$\cdot$s$^{-1}$$\cdot$0.01\%BW$^{-1}$ and a broad accessible photon energy range of $h\nu$=300-1600~eV. The temperature of the sample was approximately 14~K for all measurements. p-polarized light was used for the core level measurements and right-circularly polarized light for the angle-integrated and angle-resolved measurements in the valence band region, as shown in Figure~\ref{Figure3}c). The combined energy resolution resulting from the beamline and the hemispherical analyzer (PHOIBOS-150) varied with photon energy from $\sim$40~meV ($h\nu$=400~eV) to $\sim$147~meV ($h\nu$=1200~eV) for XPS measurements. The corresponding resolution for the valence band spectroscopy with $h\nu$=700~eV was <120~meV. Information about the sample preparation can be found in the SI section~S8.\

\subsection{Details about the electronic structure calculations}
Calculations were performed with the Korringa-Kohn-Rostoker (KKR) multiple scattering method as implemented in the SPR-KKR package \cite{Ebert2011}. This Green's function based approach allowed a fully relativistic treatment of the BTON system with varying anionic composition. First-principles calculations in the frame of density functional theory (DFT) were executed using the modified Becke Johnson (mBJ) exchange-correlation functional \cite{Tran2009,CamargoMartinez2012}, which is a meta generalized gradient approximation (meta-GGA) potential suitable for perovskite-type structures and the description of optical properties in semiconductors \cite{Niklaus2019}. The self-consistent-field (SCF) potential of the BTON systems were obtained in the atomic-sphere approximation (ASA) mode as an appropriate description for perovskite-type correlated oxide systems \cite{Lev2015}. The cubic lattice parameter a=4.113~\textup{~\AA} (ICSD \#258747) was used for all computations \cite{Hibino2017}. The angular-momentum cutoff was set to 4 and the magnetic effective field to zero. Including spin-orbit coupling, we obtained band hybridization exemplified by Ta 5\textit{d} states close to the conduction band minimum (for \ch{BaTaO_2N}) at $\Gamma$ and O 2\textit{p} states close to 3.5~eV at the $R$ point. We ensured that the convergence parameters for the KKR structure constant matrix match the converging solution of systematic, cluster-size-dependent studies on the structure constant matrix for KKR computations in tight binding mode. Atomic disorder of the anionic sites, either occupied by oxygen, nitrogen, or vacancies, was implemented within the coherent potential approximation (CPA) in SPR-KKR \cite{Ebert2011}.\

The Bloch spectral function (BSF), as the Fourier transform of the retarded Green's function in real space, allowed understanding of the characteristic oxynitride band dispersion through high-symmetry points of the simple cubic Brillouin zone (see Figure~\ref{Figure3}a) and SI Figure~S6). The BSF can be seen as the $\mathbf{k}$-resolved ground state density of states whereby the imaginary energies in the description of the BSF cause the broadening of bands \cite{Ebert2011,Minar2014}. The partial density of states used for Figure~\ref{Figure3}b) was obtained by integration of over 500~$\mathbf{k}$-points. We discuss and justify our computational approach in greater detail in the SI section~S3.\

%%%%%%%%%%%%%%%%%%%%%%%%%%%%%%%%%%%%%%%%%%%%%%%%%%%%%%%%%%%%%%%%%%%%%
\begin{acknowledgement}
The authors thank the Paul Scherrer Institute financial support as part of the PSI CROSS initiative. This work was also supported by the project Quantum materials for applications in sustainable technologies (QM4ST), funded as project No. CZ.02.01.01\slash00\slash22\slash008\slash0004572 by P JAK, call Excellent Research. The authors also thank the European Regional Development Fund and the State of Brandenburg for the Themis Z TEM (part of PISA). The SX-ARPES staff at the Swiss Light Source thanks Leonard Nue for exceptional technical support. A. H. appreciates guidance using the SPR-KKR package from Laurent Nicolaï and fruitful discussions with Christof Schneider and Jan Bosse.\
\end{acknowledgement}
%%%%%%%%%%%%%%%%%%%%%%%%%%%%%%%%%%%%%%%%%%%%%%%%%%%%%%%%%%%%%%%%%%%%%
\begin{suppinfo}
The following files are available free of charge.
\begin{itemize}
  \item The Supporting Information ($*$.pdf) comprises further sample characterization by means of x-ray diffraction, TEM, and PES, explanations concerning the computational approach, the PES data processing as well as the PES sample preparation, and the derivation of the model for nonuniformly doped oxynitrides.
  \item Additional data consisting of PES raw data files in $*$.h5 format, the results from RBS/ERDA measurements, and the quantification of the N:O ratio using TEM, as shown in the SI Figure~S4, are summarized at the Zenodo open data repository accessible via \href{https://doi.org/10.5281/zenodo.13350672}{https://doi.org/10.5281/zenodo.13350672}.
\end{itemize}
\end{suppinfo}
%%%%%%%%%%%%%%%%%%%%%%%%%%%%%%%%%%%%%%%%%%%%%%%%%%%%%%%%%%%%%%%%%%%%%
\bibliography{BTON_MS}
%%%%%%%%%%%%%%%%%%%%%%%%%%%%%%%%%%%%%%%%%%%%%%%%%%%%%%%%%%%%%%%%%%%%%
\end{document}